\documentclass{ws-procs9x6}
\begin{document}

\title{A Semiclassical Approach to Fusion Reactions\footnote{
\uppercase{T}his work was supported in part by the Brazilian
agencies \uppercase{CNP}q, \uppercase{FAPESP},
\uppercase{FAPERJ} and the \uppercase{I}nstituto do \uppercase{M}il\^enio de
\uppercase{I}nforma\c c\~ao \uppercase{Q}u\^antica-\uppercase{MCT},
\uppercase{B}razil.}}
\author{M.~S. Hussein}

\address{Instituto de F\'\i sica, Universidade de S\~ao Paulo \\
CP 66318, 05389-970, S\~ao Paulo SP, Brazil \\
E-mail: hussein@fma.if.usp.br}

\author{L.~F. Canto and R.~Donangelo}
\address{Instituto de F\'\i sica, Universidade Federal do Rio de Janeiro,\\
CP 68528, 21941-972, Rio de Janeiro RJ, Brazil \\
E-mail: canto@if.ufrj.br, donangelo@if.ufrj.br}

\maketitle

\abstracts{
The semiclassical method of Alder and Winther is generalized to study
fusion reactions. As an illustration, we evaluate the fusion cross section
in a schematic two-channel calculation. The results are shown to be
in good agreement with those obtained with a quantal Coupled-Channels
calculation. We suggest that in the case of coupling to continuum
states this approach may provide a simpler alternative to the Continuum
Discretized Coupled-Channels method.}

\section{Introduction}

The importance of Coupled-Channels effects on the fusion cross section has
been investigated by several authors\cite{Da85}. These studies have 
established that the main effect of the coupling of the entrance channel with
other bound channels is to produce a pronounced enhancement of the fusion
cross section at sub-barrier energies. A more complicated situation arises
when the reaction involves weakly bound nuclei. In such cases, the elastic
channel is strongly coupled with the breakup channel, which corresponds to
states of three or more bodies in the continuum. The total fusion cross
section, $\sigma_{F}$, is then the sum of different processes: the complete
fusion cross section, $\sigma_{CF}$, where all projectile's and target's
nucleons merge into a compound system, and incomplete fusion cross sections,
$\sigma_{ICFi}$, where only the $i^{th}$ fragment of the projectile fuses with 
the target while the remaining ones come out of the interaction region.

A recent review of the experimental and theoretical work on the fusion of
unstable or weakly bound nuclei can be found in ref.\cite{Hu03}. The first
theoretical studies\cite{Hu92,Ta93,Da94} used schematic models, which stress 
particular aspects of the fusion process. More recently, sophisticated quantum
Coupled-Channels calculations have been performed\cite{Ha00,Di02}.
These calculations approximate the continuum by a discrete set of states,
according to the Continuum Discretized Coupled-Channels method (CDCC). Those
calculations led to the conclusion that in collisions with very heavy targets
the coupling to the continuum has a strong influence on the complete fusion
cross section, The progress in the experimental study of these
collision is more recent, since only recently unstable beams at barrier
energies became available\cite{Ko98,Re98,Si99,Tr00}. Besides, measurements at
sub-barrier energies are very hard to perform, owing to the low intensity of
the unstable beams. Although some recent measurements of the fusion cross
section in collisions of unstable beams from heavy targets show an enhancement
at sub-barrier energies\cite{Ko98,Tr00}, more data are needed for a final
conclusion. On the other hand, data on the fusion cross section in reactions
induced by light weakly bound stable projectiles have been available for a
longer time\cite{Hi02}.

The importance of the details of the CDCC basis in calculations of the fusion
cross section, pointed out in ref.\cite{Di02}, indicates
that a simple approximation for the breakup channel can only be used for very
qualitative estimates, like that of ref.\cite{Ca03}.
A reasonable alternative is the use of the semiclassical method of 
Alder-Winther (AW)\cite{AW}.  This method was originally proposed to study
Coulomb excitation of collective states and it was latter generalized to other
nuclear reactions, including the excitation of the breakup channel\cite{AW1}.
More recently, it has been used to study the breakup of $^{8}$B in the
$^{8}$B + $^{58}$Ni collision\cite{Ma01} for a comparison with the
CDCC calculations of Nunes and Thompson\cite{NT98}. The discretization of the
continuum space was carried out in the same way as in ref.\cite{NT98} and the results
obtained with the AW approximation were shown to be in good agreement with
those of the CDCC method. In the present work, we show how the AW method can
be used to evaluate the complete fusion cross section in collisions of weakly
bound projectiles and discuss its validity in a schematic two-channel example.

This paper is organized as follows: in section 2 we introduce the 
Alder-Winther method and show how it can be used to evaluate the complete fusion
cross section. An application to a schematic model that mimics the $^{6}$He +
$^{238}$U is made. In section 3 we present the conclusions of this work.

\section{The Alder- Winther method.}

Let us consider a collision described by the projectile-target separation
vector, $\mathbf{r}$, and the relevant intrinsic degrees of freedom of the
projectile, represented by $\xi$. For simplicity, we neglect the internal
structure of the target. The projectile's Hamiltonian is
\begin{equation}
h=h_{0}(\xi)+\mathcal{V}(\mathbf{r},\xi),\label{h}
\end{equation}
where $h_{0}(\xi)$ is the intrinsic Hamiltonian and $V(\mathbf{r},\xi)$
represents the projectile-target interaction. The eigenvectors of 
$h_{0}(\xi)$ are given by 
\begin{equation}
h_{0}~\left\vert \phi_{\alpha}\right\rangle =\varepsilon_{\alpha}~\left\vert
\phi_{\alpha}\right\rangle .\label{av}
\end{equation}
The Alder- Winther method is implemented as follows. First, one uses
classical mechanics for the variable $\mathbf{r}$. In its original version, a
Rutherford trajectory $\mathbf{r}_{l}(t)$ was used. The trajectory depends on
the collision energy, $E,$ and the angular momentum, $l$. In our case, we use
the solution of the classical equation of motion with the potential
$V(\mathbf{r)=}$\ $\left\langle \phi_{0}\right\vert \mathcal{V}(\mathbf{r}
,\xi)\left\vert \phi_{0}\right\rangle ,$ where $\left\vert \phi_{0}
\right\rangle $\ is the ground state of the projectile. Using the trajectory,
the coupling interaction becomes a time-dependent interaction in the $\xi
$-space. \ That is, $\mathcal{V}(\xi,t)\equiv\mathcal{V}(\mathbf{r}_{l}
(t),\xi).$ Then the dynamics in the intrinsic space is treated as a 
time-dependent quantum mechanics problem, according to the
Schr\"{o}dinger equation
\begin{equation}
h~\psi(\xi,t)=\left[  h_{0}(\xi)+\mathcal{V}(\xi ,t)\right]  ~\psi
(\xi,t)=i\hbar~\frac{\partial\psi(\xi,t)}{\partial t}.\label{Sch}
\end{equation}
Expanding the wave function in the basis of intrinsic eigenstates,
\begin{equation}
\psi(\xi,t)=\sum_{\alpha}a_{\alpha}(l,t)~\phi_{\alpha}(\xi)~e^{-i\varepsilon
_{\alpha}t/\hbar},\label{exp}
\end{equation}
and inserting the expansion in eq.(\ref{Sch}), one obtains the Alder-
Winther equations
\begin{equation}
i\hbar~\dot{a}_{\alpha}(l,t)=\sum_{\beta}~\left\langle \phi_{\alpha
}\right\vert \mathcal{V}(\xi,t)\left\vert \phi_{\beta}\right\rangle
~e^{i\left(  \varepsilon_{\alpha}-\varepsilon_{\beta}\right)  t/\hbar
}~a_{\beta}(l,t).\label{AW}
\end{equation}
These equations should be solved with initial conditions $a_{\alpha
}(l,t\rightarrow-\infty)=\delta_{\alpha0},$ which means that before the
collision ($t\rightarrow-\infty$) the projectile was in its ground state. The
final population of channel $\alpha$ in a collision with angular momentum
$l$\ is $P_{l}(\alpha)=\left\vert a_{\alpha}(l,t\rightarrow+\infty)\right\vert
^{2}$ and the cross section is
\begin{equation}
\sigma_{\alpha}=\frac{\pi}{k^{2}}~\sum_{l}\left(  2l+1\right)  ~P_{l}
(\alpha).\label{PW1}
\end{equation}
A similar procedure can be used to derive angular distributions.

The AW method can be extended to evaluate the fusion cross section as
follows. The starting point is the general expression for the fusion
cross section in multi-channel scattering
\begin{equation}
\sigma_{F}=\sum_{\alpha}~\sigma_{F}^{(\alpha)};\;\;\;\sigma_{F}^{(\alpha
)}=\frac{\pi}{k^{2}}\sum_{l}\left(  2l+1\right)  ~P_{l}^{F}(\alpha),
\label{sigfus}
\end{equation}
with the fusion probability for the $l^{th}$-partial-wave in channel $\alpha$
given by
\begin{equation}
P_{l}^{F}(\alpha)=\frac{4k}{E}~\int dr~\left\vert u_{l}(k,r)\right\vert
^{2}~W_{\alpha}^{F}(r). \label{PQM}
\end{equation}
Above, $u_{l}(k,r)$ represents the radial wave function for the $l^{th}
$-partial-wave in channel $\alpha$ and $W_{\alpha}^{F}$ is the absolute value
of the imaginary part of the optical potential in this channel arising from
fusion. To use the AW method to evaluate the fusion cross section, we make the
approximation
\begin{equation}
P_{l}^{F}(\alpha)\simeq T_{l}~\left\vert a_{\alpha}(l,t_{ca})\right\vert ^{2}.
\label{PLAW}
\end{equation}
In the above equation, $T_{l}$ is the tunneling probability and $\left\vert
a_{\alpha}(l,t_{ca})\right\vert ^{2}$ is the probability that the projectile
is in the state $\left\vert \phi_{\alpha}\right\rangle $ when the system
reaches closest approach.

\begin{figure}[th]
\centerline{\epsfxsize=3.9in\epsfbox{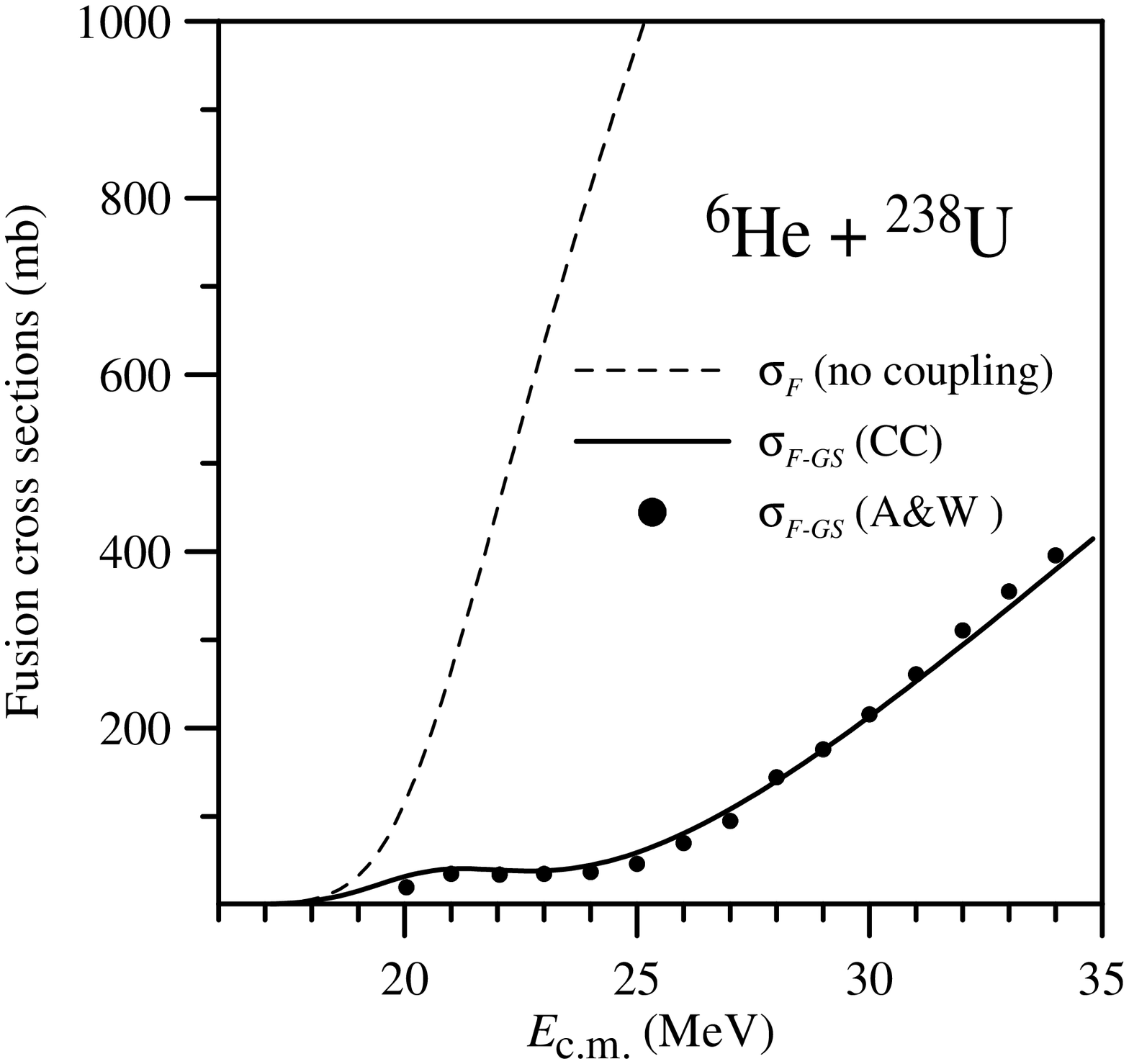}}\caption{Contribution from the
elastic channel to the fusion cross section. Results obtained with the
Alder- Winther approximation are represented by solid circles. For comparison,
results obtained with a coupled-channel calculation (solid line) and without
channel-coupling (dashed line) are also shown. }
\label{inter}
\end{figure}

\bigskip

We have performed a preliminary calculation for a two-channel case, studying
the scattering of $^{6}$He projectiles on a $^{238}$U target, at near barrier
energies. The weakly bound $^{6}\mathrm{He}$\ nucleus dissociates into
$^{4}\mathrm{He}$\ and two neutrons, with threshold energy $B=0.975$ MeV. The
elastic channel is strongly coupled to the breakup channel and this
coupling has a strong influence on the fusion cross sections. Here
we represent the breakup channel by a single effective channel and evaluate
the complete fusion cross section using the semiclassical method mentioned
above. In this approximation, the complete fusion cross section
corresponds to the contribution from the elastic channel to eq.(\ref{sigfus}).
For simplicity, we neglect the excitation energy and the spin of the effective
channel. We adopt a form factor with the radial dependence of the electric
dipole coupling and the strength is chosen arbitrarily, in such a way that the
coupling modifies the cross section predicted by the one dimension penetration
barrier appreciably. In figure 1, we compare results obtained with the AW
method with those of a coupled channel calculation and also with those
obtained with the neglect of channel coupling. We adopt Wood-Saxon shapes for
the real and imaginary potentials, with the parameters: $V_{0}=-50~MeV,$
$r_{0r}=1.25~fm,~a_{r}=0.65~fm,$ $W_{0}=-50~MeV,~r_{0i}=1.0~fm,~a_{i}
=0.65~fm.$ We conclude that the semiclassical results are very close to those
of a full coupled-channel calculation. It should be remarked, however, that
this good agreement does not occur at sub-barrier energies. In this energy
range the classical trajectory does not reach the barrier radius and therefore
the effective barrier lowering that enhances the cross section is not
accounted for.

\section{Conclusions}

The semiclassical Coupled-Channels theory of fusion reactions presented here
is a natural extension of what has been done for other reaction channels. As
it has been shown in a previous study of the breakup cross section\cite{Ma01},
it allows a realistic description of the breakup channel, including
continuum-continuum coupling. Although the calculation presented was restricted
to a schematic model two-channel model, an extension to a large set of
continuum states along the lines of ref.\cite{Ma01} should present no major
difficulties.

It should be pointed out that the present model can be extended to calculate
the fusion of the fragment that contains all or most of the charge of the
projectile. In a way, it could be considered an improved semi-quantal version
of the classical three-body model of Hinde \textit{et al}.\cite{Hi02}. Work
is in progress to accommodate both complete and incomplete 
fusion in the theory and thus supplying a simplified albeit accurate version
of the CDCC.

\end{document}